\documentstyle{prhep97}
\input psfig
\newcommand{\ds}{\displaystyle}
\makeatletter
\let\chapter\hid@chapter
\makeatother
\begin{document}
%\pagenumbering{empty}

% The following definitions need to be customised;

% Will appear on page headings
\authorrunning{R.\,Devenish}
\titlerunning{{\talknumber}: Structure Functions}
 
% Now the full name of author and talk

% For plenary talks, the talk number is that of the session
\def\talknumber{PL3} 

\title{{\talknumber}: Structure Functions\thanks{Extended version
of the writeup of the plenary talk given at the EPS HEP97 Conference,
Jerusalem Aug. 1997.}\hfil OUNP-98-01 }
\author{Robin\,Devenish
(r.devenish@physics.ox.ac.uk)}
\institute{ Physics Dept., Oxford University, UK}

\maketitle

\begin{abstract}
Recent measurements of unpolarised and polarised nucleon structure
functions and $F_2^{\gamma}$ are reviewed. The implications for
QCD and the gluon momentum distribution are discussed.
The status of the understanding of $\sigma_{tot}^{\gamma^*p}$ in
the transition region between real photoproduction and deep-inelastic
scattering is summarised briefly. 
\end{abstract}
\section{Introduction}
\label{sec:intro}

This talk 
covers three areas: unpolarised deep-inelastic scattering 
(DIS) data, parton distributions and
associated phenomenology (Sec.~\ref{sec:unpol}); 
nucleon spin structure (Sec.~\ref{sec:spin}) and the status of
$F_2^\gamma$ measurements (Sec.~\ref{sec:f2g}).
New measurements from the Tevatron relevant for parton determination
such as W decay asymmetries, Drell-Yan asymmetries, direct $\gamma$
and inclusive jet cross-sections are covered by Weerts~\cite{weerts}. 
Diffractive DIS and the diffractive structure
function are covered by Eichler~\cite{eichler},
recent measurements of $\alpha_S$ by 
Ward~\cite{ward} and the status of DIS measurements at very large $Q^2$
from HERA are summarised by Elsen~\cite{elsen}.

\section{Unpolarised Deep Inelastic Scattering }
\label{sec:unpol}

\begin{figure}[ht]
\vspace*{13pt}
\begin{center}
\psfig{figure=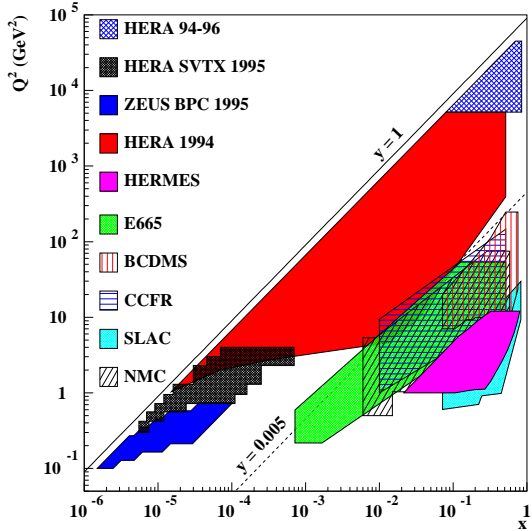,height=7cm} 
\caption{Regions in the ($x,Q^2$) plane for fixed target and collider
DIS experiments.}
\label{fig:kinreg}
\end{center}
\end{figure}

The kinematic variables describing
DIS are $\ds Q^2=-(k-k^\prime)^2,~x=Q^2/(2p.q)$, $y=(p.q)/(p.k)$,
where $q=k-k^\prime$ and $k$, $k^\prime$, $p$ are the 4-momenta
of the initial and final lepton and target nucleon respectively. At
fixed $s$, where $\ds s=(k+p)^2$, and ignoring masses the variables
are related by $\ds Q^2=sxy$. 
The expression for the double differential
neutral-current DIS cross-section is
{\small \begin{equation}
\frac {d^2\sigma (l^{\pm}N) } {dxdQ^2} =  \frac {2\pi\alpha^2} {Q^4 x}  
\left[Y_+\,F_2(x,Q^2) - y^2 \,F_L(x,Q^2)
\mp Y_-\, xF_3(x,Q^2) \right],
\label{eq:NCxsec}
\end{equation}} 
where $\displaystyle Y_\pm=1\pm(1-y)^2$ and $F_i~(i=2,3,L)$ are the
nucleon structure functions. 
For $Q^2$ values much below that of the $Z^0$ mass squared, the parity 
violating structure function $xF_3$ is 
negligible. $F_L$ is a significant contribution only at large $y$. At
HERA both $F_3$ and $F_L$ are treated as calculated corrections and 
the $F_2$ data quoted is that corresponding to $\gamma^*$ exchange only.
The kinematic coverage of recent fixed target and HERA collider 
experiments is shown in Fig.~\ref{fig:kinreg} and more details are
given in Table~\ref{tab:datsum}.

\begin{table}[htb]
\caption{Summary of recent structure function experiments.
All the data referred in this table are available from 
the Durham HEPDATA database, at 
http://durpdg.dur.ac.uk/HEPDATA on the world wide web.
}
\centerline{
\begin{tabular}{llllcccc}\\
 \hline
 Beam(s)~~ & Targets~~& Experiment~~  & $Q^2$ (GeV$^2$) & 
 $x$ & $R$~~ & Status \\
 \hline
 $e^-$ & p,d,A & SLAC  & $0.6-30$ & $0.06-0.9$ & yes & complete \\
 $\mu$ & p,d,A & BCDMS  & $7-260$ & $0.06-0.8$ & yes & complete \\
 $\mu$ & p,d,A & NMC  & $0.5-75$ & $0.0045-0.6$ & yes & complete \\
 $\mu$ & p,d,A & E665  & $0.2-75$ & $8\cdot10^{-4}-0.6$ & no & 
 compete \\ 
 $\nu,\bar{\nu}$ & Fe &CCFR & $1.-500.$ & $0.015-0.65$ & yes &
 complete \\
 $e^\pm$, p& - & H1  & $0.35-5000$ & ~~$6\cdot10^{-6}-0.32$ & 
~~estimate &  running \\
 $e^\pm$, p& - & ZEUS  & $0.16-5000$ & $~3\cdot10^{-6}-0.5~$ & no & 
 running\\
\hline\\
\end{tabular}}
\label{tab:datsum}
\end{table}

The vast bulk of nucleon structure function data is for $F_2$ and here 
the overall situation is rather pleasing. The fixed target programme
is complete with the publication in the last 18 months of the final
data from NMC~\cite{nmcf2} and E665~\cite{e665} to add to the older
data from SLAC and BCDMS that still play an important role in global
fits to determine parton distribution functions (PDFs).
\begin{figure}[tbp]
\vspace*{13pt}
\begin{center}
\psfig{figure=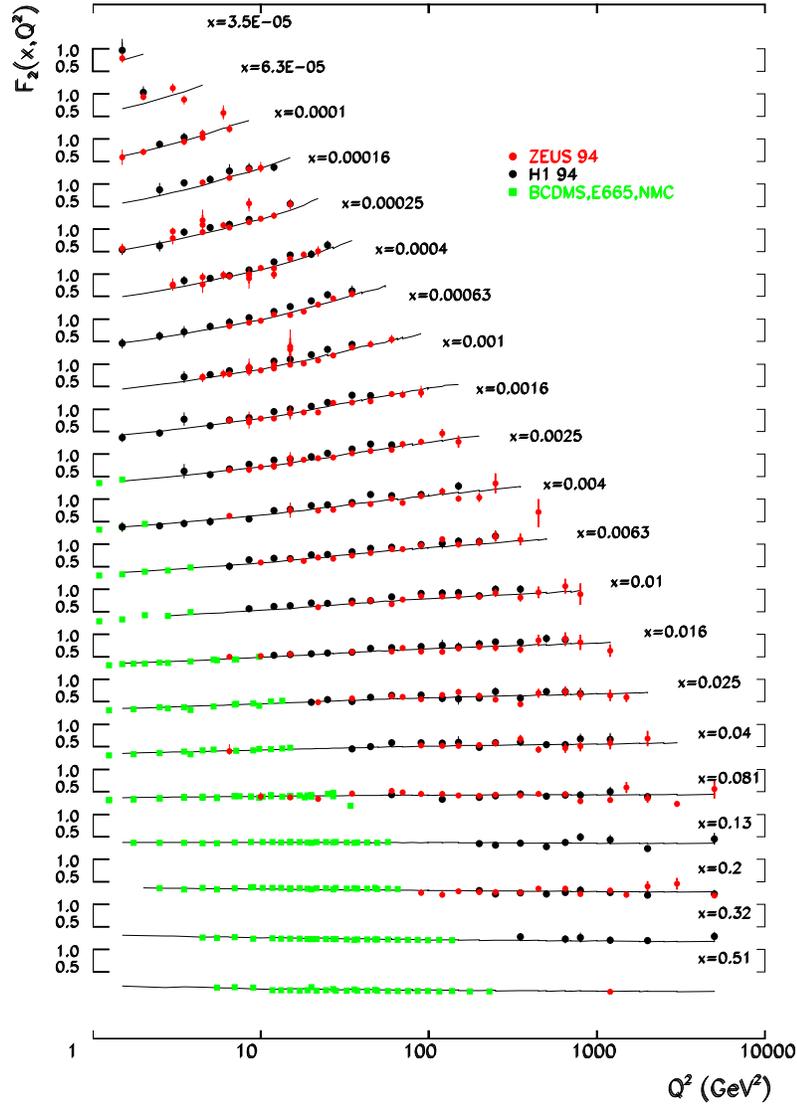,bbllx=0pt,bblly=0pt,bburx=570pt,bbury=840pt,height=0.9\textheight} 
\caption{$F_2^p$ data from HERA(94) and fixed target experiments at fixed
$x$ as a function of $Q^2$. The curves shown are the NLO DGLAP 
QCD fit used to smooth the data during unfolding.}
\label{fig:zall_q2}
\end{center}
\end{figure}
The first high statistics  data from
the 1994 HERA run were published by H1~\cite{h1_94f2} and 
ZEUS~\cite{z_94f2} last year. The $F_2$ data now covering  4 decades in $Q^2$
and 5 decades in $x$ are summarised in Fig.~\ref{fig:zall_q2}. The data
from the fixed target and HERA collider experiments are consistent
with each other in shape and normalisation and show the pattern of
scaling violations expected from perturbative QCD (pQCD).
The systematic errors for the fixed target experiments are typically 
less than 5\% and those for H1 and ZEUS around 5\% for $Q^2<100\,$GeV$^2$, 
above this value the errors become statistics dominated. 
Fairly recently CCFR published an update of their high statistics
$F_2^{\nu Fe}$ and $xF_3^{\nu Fe}$ data~\cite{ccfr}, following an
improved determination of energy calibrations. The CCFR data and
a determination of $\alpha_S$ are described in more
detail by De Barbaro~\cite{barbaro}.
\begin{figure}[tbp]
\vspace*{13pt}
\begin{tabular}[ht]{ll}
\psfig{figure=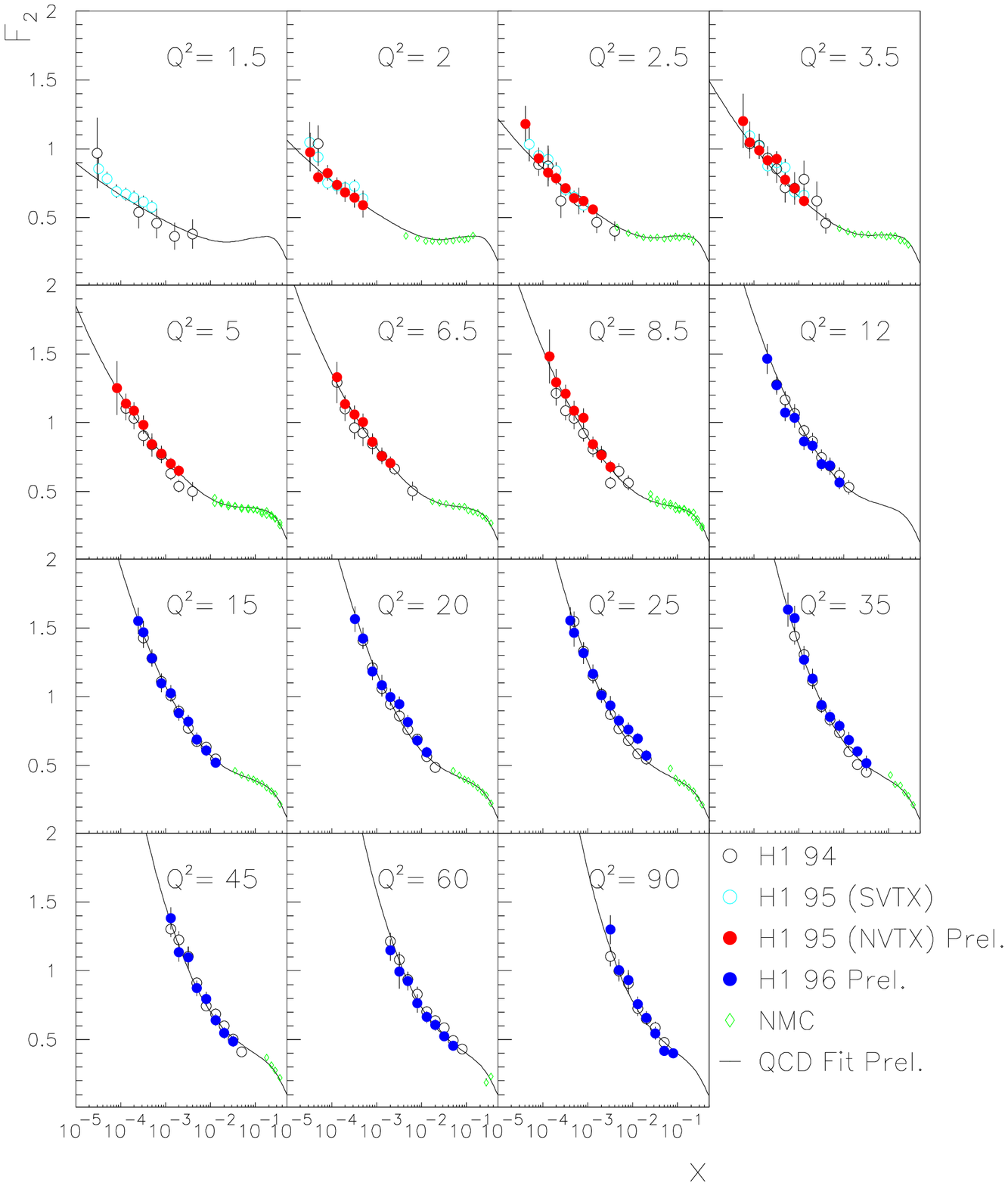,width=.45\textwidth}&
\psfig{figure=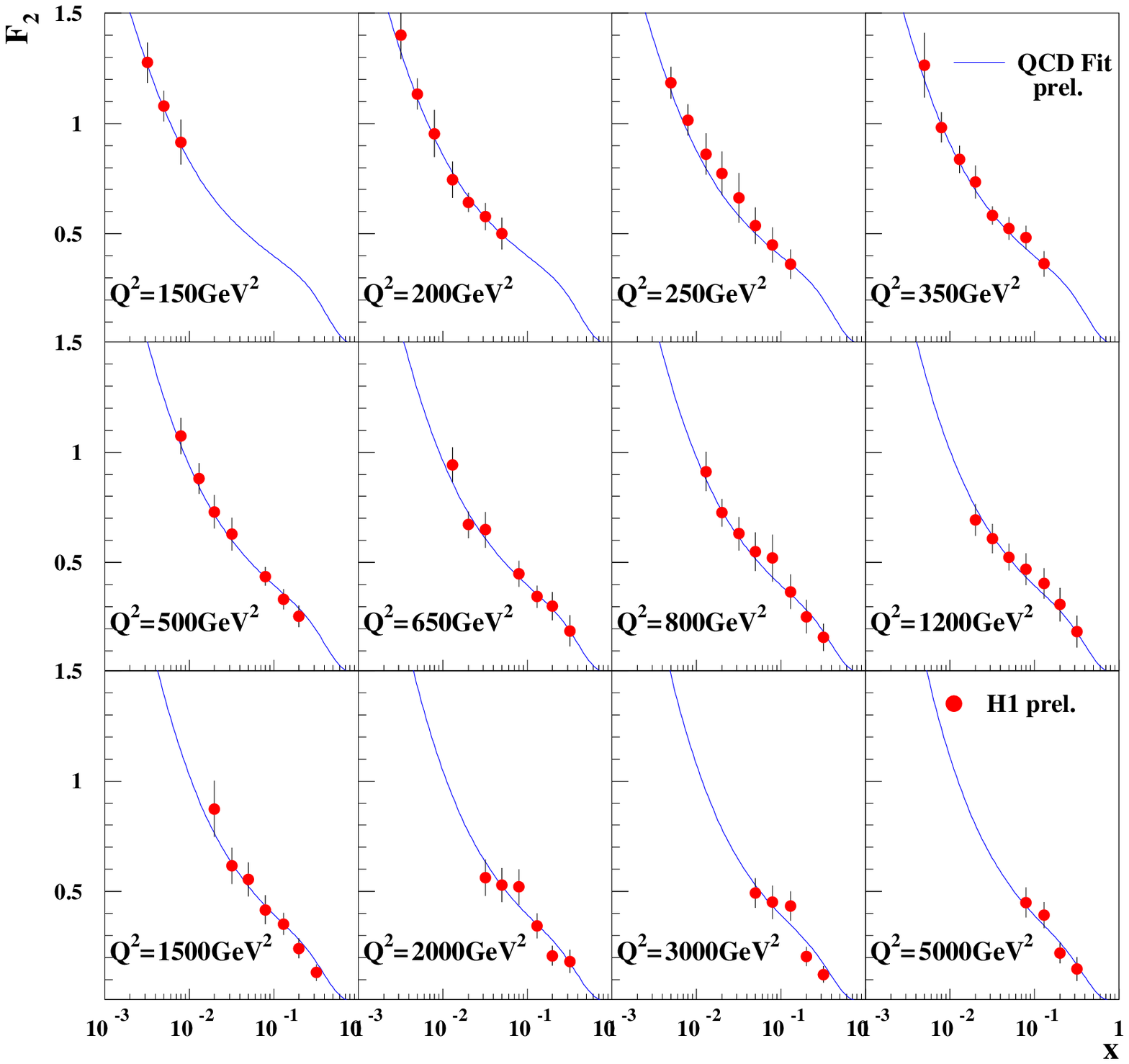,width=.45\textwidth} 
\end{tabular}
\caption{Preliminary H1 $F_2$ data: left $1<Q^2<100\,$GeV$^2$ from
HERA 1995/6; right $150<Q^2<5000\,$GeV$^2$ from
HERA 1995-97 runs. The curves show the $Q^2$ evolution of a NLO QCD
fit to H1 data on the left and evolved to higher $Q^2$. }
\label{fig:h1f2q2}
\end{figure}

H1 has submitted preliminary values of $F_2$ from more recent HERA runs.
The data are shown in Fig.~\ref{fig:h1f2q2}: 
on the left for $1<Q^2<100\,$GeV$^2$
(the region covered by the improved H1 rear
detector) is from $5.4\,$pb$^{-1}$ taken in 1995/96~\cite{h1_9596f2};
on the right for $150<Q^2<5000\,$GeV$^2$ is from $22\,$pb$^{-1}$ 
accumulated over the period 1995-97~\cite{bassler}. 
Also shown in Fig.~\ref{fig:h1f2q2} is a NLO QCD fit to H1 data with
$Q^2<120\,$GeV$^2$, which is evolved to cover the region
of the higher $Q^2$ data. All the new data are
well described by the QCD curves and the characteristic steep rise
of $F_2$ as $x$ decreases is seen up to the largest $Q^2$ values. 
 
\subsection{The low $Q^2$ transition region}

\begin{figure}[htbp]
\vspace*{13pt}
\begin{center}
\psfig{figure=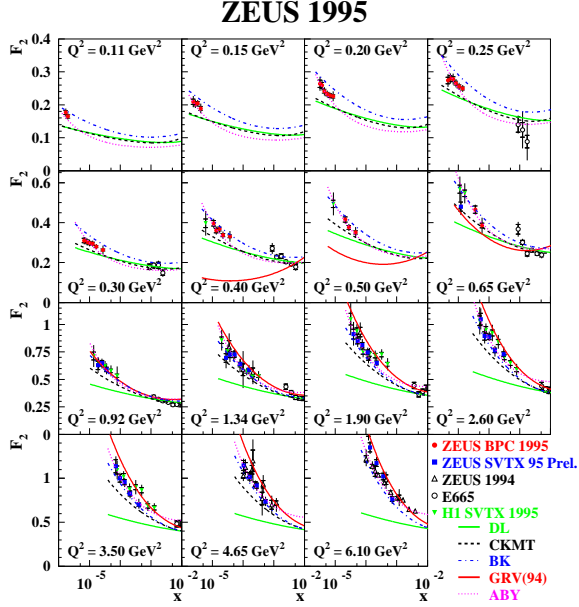,height=8cm} 
\caption{$F_2$ data from E665, H1 and ZEUS at very small values of
$Q^2$. The curves are described in the text.}
\label{fig:f2_loq2}
\end{center}
\end{figure}

One of the surprises of the HERA $F_2$ data is the low scale from 
which NLO QCD evolution seems to work. H1 and ZEUS have now measured 
the cross-sections and hence $F_2$ from the safely DIS at $Q^2\sim
6\,$GeV$^2$ through the transition region to $Q^2=0$, using
a combination of new detectors very close to the electron beam
line and by shifting the primary interaction vertex in the proton
direction by $70\,$cm. The 
data are shown in Fig.~\ref{fig:f2_loq2} and are described in
detail in refs.~\cite{h1svx95,zbpc95,p646}. Also shown in the
figure are data from the E665 experiment~\cite{e665} which had
a special trigger to allow measurements at small $x$ and $Q^2$.
As $Q^2\to 0$ $F_2$ must tend to zero at least as fast as $Q^2$,
it is often more convenient to consider
\begin{equation}
\sigma_{tot}^{\gamma^*p}(W^2,Q^2)\approx 
{4\pi^2\alpha\over Q^2}F_2(x,Q^2)
\end{equation}
which is valid for small $x$ and where $W^2\approx Q^2/x$ is the 
centre-of-mass energy squared of the $\gamma^* p$ system. 
For $Q^2>1\,$GeV$^2$, the steep rise of $F_2$ as $x$ decreases is
reflected in a steeper rise of $\sigma_{tot}^{\gamma^*p}$ with $W^2$
than the slow increase shown by $\sigma_{tot}^{\gamma p}$ and
characteristic of hadron-hadron total cross-sections.

Two very different approaches, both proposed before the HERA
measurements, may be taken as paradigms. Gl\"uck, Reya and Vogt (GRV)
\cite{grv1} have long advocated a very low starting scale as part
of their approach to generate PDFs `dynamically' using NLO QCD. 
Predictions from their most recent parameterisation~\cite{grv2} 
are shown as the black solid line in Fig.~\ref{fig:f2_loq2},
starting in the $Q^2=0.4\,$GeV$^2$ bin and upwards. The data
are in reasonable agreement with the theory down as far as the
$Q^2=0.92\,$GeV$^2$ bin. The other approach, that of Donnachie
and Landshoff (DL)~\cite{donlan}, is an extension of Regge 
parameterisations that describe hadron-hadron and real photoproduction
data well. The form that DL use to describe $\sigma_{tot}^{\gamma^*p}$ 
is
\begin{equation}
\sigma_{DL}=A(Q^2)(W^2)^{\alpha_P-1}+B(Q^2)(W^2)^{\alpha_R-1},
\end{equation}
where $\alpha_P$ and $\alpha_R$ are the intercepts of the 
Pomeron and Reggeon trajectories respectively with values
$\alpha_P=1.08$, $\alpha_R=0.05$, determined from hadron-hadron data.
The DL model gives the trend of the energy dependence 
of the very low $Q^2$ $\sigma_{tot}^{\gamma^*p}$ HERA data, 
up to $Q^2\sim 0.4\,$GeV$^2$,
though the normalisation of the model is a bit on the low side. The
DL curves in Fig.~\ref{fig:f2_loq2} are the solid grey lines. 
In~\cite{p647} the ZEUS collaboration has investigated the
transition region. From NLO QCD fits with starting scales of 
$Q^2_0=0.4,~0.8,~1.2\,$GeV$^2$ it is found that only the latter two
give acceptable descriptions of the data. For the limit
$Q^2\to 0$ a DL form is used. From these two approaches the
transition to pQCD occurs in the $Q^2$ range $0.8-1.2\,$GeV$^2$. 

The advent of accurate data from HERA has 
prompted many groups to try to model the behaviour of 
$\sigma_{tot}^{\gamma^*p}$ throughout the transition region. 
Very briefly: the model of Capella et al
(CKMT)~\cite{ckmt} uses a DL form at low $Q^2$ but allows the Regge 
intercepts $\alpha_P,~\alpha_R$ to become $Q^2$ dependent, for 
$Q^2>2\,$GeV$^2$ DGLAP evolution gives the $Q^2$ dependence; 
Abramowicz et al (ALLM)~\cite{allm} follow a similar approach 
but use a QCD inspired parameterisation at large $Q^2$; Badelek \&
Kwiecinski (BK)~\cite{bk} take 
$\ds F_2=F_2^{VMD}+Q^2F_2^{QCD}/(Q^2+Q_0^2)$ where $F_2^{VMD}$ is given
by strict $\rho,~\omega,~\phi$ VMD and the QCD scale parameter $Q_0^2$ is
chosen to be $1.2\,$GeV$^2$; Schildknecht and Spiesberger 
(ScSp)~\cite{scsp} revive the idea of GVMD to fit data for $0<x<0.05$
and $0<Q^2<350\,$GeV$^2$; Kerley and Shaw~\cite{ks} modify the idea
of long-lived hadronic fluctuations of the photon to include jet
production; Gotsman, Levin \& Maor~\cite{glm} also follow this approach 
but have an additional hard QCD term; finally Adel, Barreiro \& 
Yndurain (ABY)~\cite{aby} have developed a model with an input $x$
dependence of the form $a+bx^{-\lambda}$ with the two terms representing
`soft' and `hard' contributions which evolve independetly with $Q^2$.
Fig.~\ref{fig:f2_loq2} shows some of these models against the low $Q^2$
data. Although most of them give a reasonable description of the trends
in $x$ and $Q^2$, only the ScSp and ABY models (which were fit to the 
data) give the details correctly. In fact these two models also have defects 
as they are not able to describe the low energy $\sigma_{tot}^{\gamma p}$
data~\cite{sheklp97}. Very recently Abramowicz and Levy~\cite{allm2}
have updated the ALLM parameterisation by including all the recent HERA
data in the fit and result gives a satisfactory description of both
the $Q^2$ and $W^2$ dependence. However, while this represents an advance,
it is still true to say that more work needs to be done before the 
low $Q^2$, low $x$ region is completely understood.
More details of many of these models
are given in the review by Badelek and Kwiecinski~\cite{bkrev}.

\subsection{QCD and parton distributions}

The striking rise of $F_2$ as $x$ decreases was at
first thought, at least by some, to be evidence for the singular behaviour
of the gluon density $xg\sim x^{-\lambda}$ with $\lambda\sim 0.3-0.5$
proposed by Balitsky et 
al~\cite{bfkl} (BFKL) and a breakdown of `conventional' pQCD as embodied
in the DGLAP equations. By the time of the EPS HEP95
conference in Brussels~\cite{eisele} the pendulum had swung 
the other way, largely through the work of Ball \& Forte~\cite{bf} on 
`double asymptotoc scaling' (DAS) and the success of GRV(94)~\cite{grv2} 
in describing the data. In both cases the rise in $F_2$ is generated
through the DGLAP kernels with a non-singular input.
 It is clear from Figs.~\ref{fig:zall_q2},
\ref{fig:h1f2q2} that NLO DGLAP $Q^2$ evolution can describe the $F_2$
data from $Q^2\approx 1.5\,$GeV$^2$ to the highest values of
$5000\,$GeV$^2$. The two global fitting teams in their most recent
determinations of the PDFs (CTEQ4~\cite{cteq4} and MRS(R)~\cite{mrsr}), 
which include the HERA 1994 data, now use starting scales of
around $1\,$GeV$^2$. The quality of the fits is good with $\chi^2/ndf$
in the range $1.06-1.33$ and it is found that the gluon distribution is
now non-singular in $x$ at the input scale with the quark sea still mildly
singular. Both CTEQ and MRS give PDFs for $\alpha_S(M_Z^2)$ in the range 
$0.113-0.120$ as there is some indication that the more recent 
determinations~\cite{ward,bethke} give a somewhat larger value than 
the `DIS value' of 0.113 determined from a fit to 
BCDMS and SLAC data~\cite{vm}. The extension of accurate measurements
to low $x$ provided by the HERA(94) data has led to a big improvement
in the knowledge of the gluon density. At low $x$ the gluon drives the
scaling violations through 
\begin{figure}[tbp]
\vspace*{13pt}
\begin{tabular}[ht]{ll}
\psfig{figure=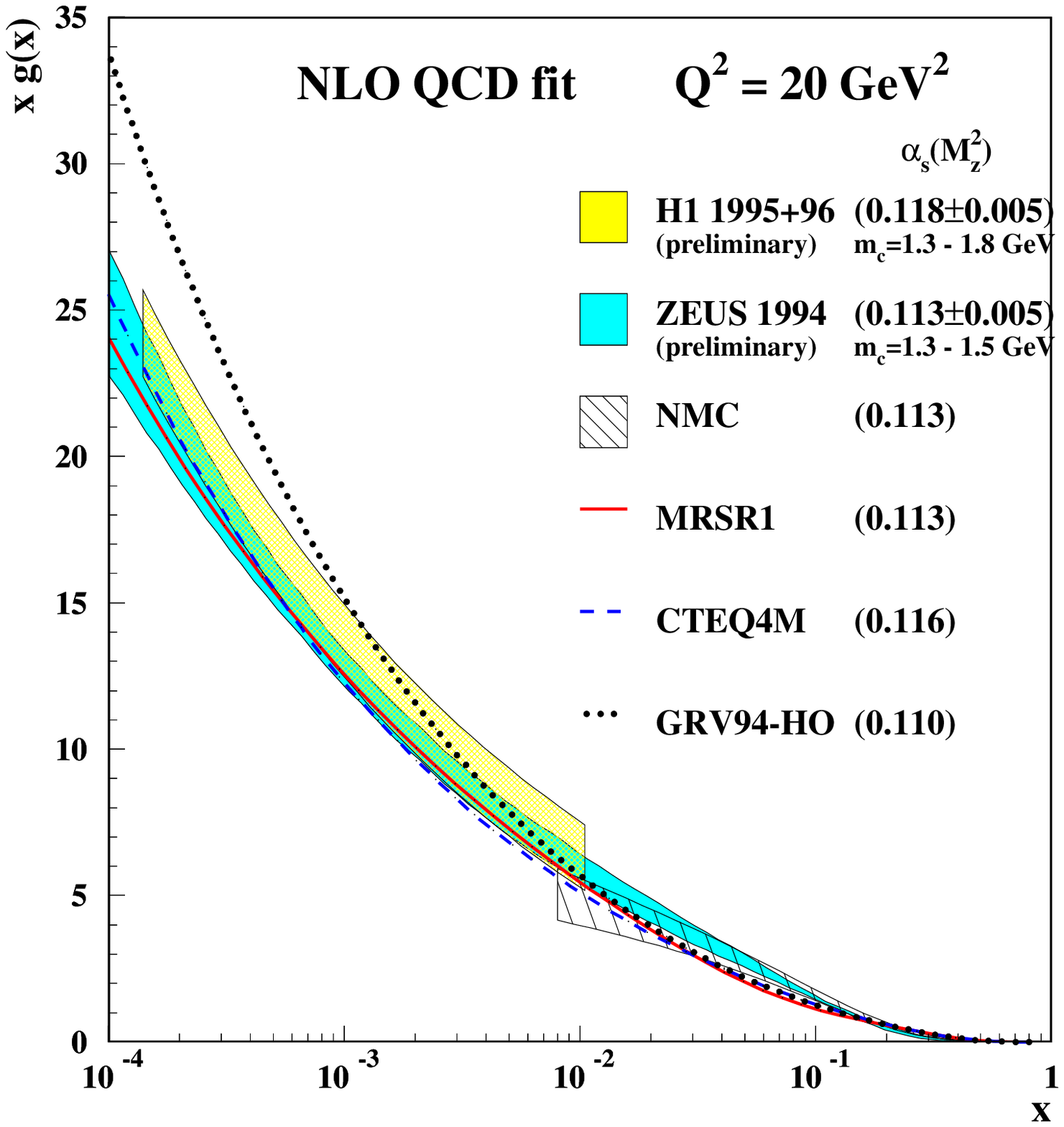,width=.45\textwidth}&
\psfig{figure=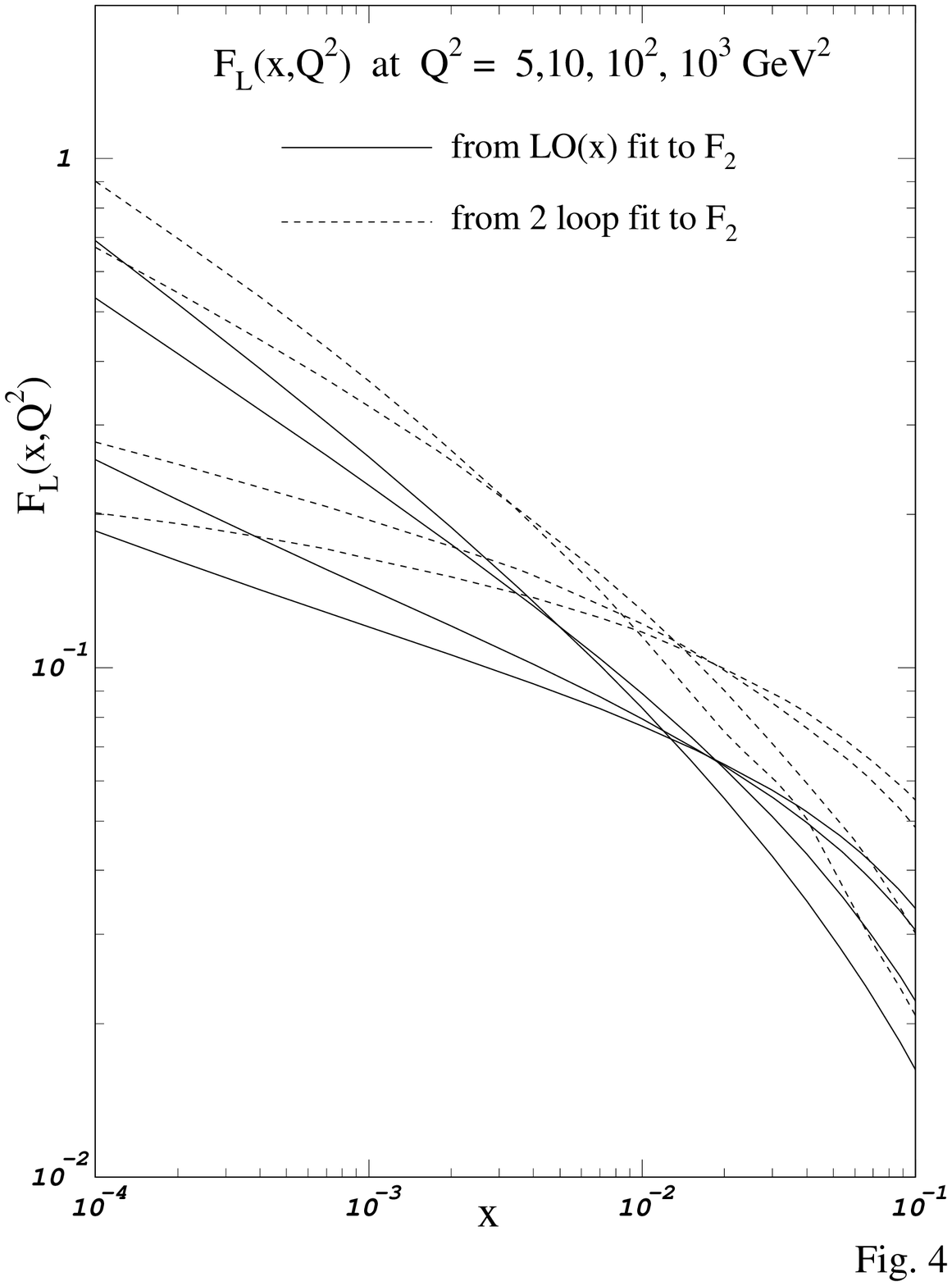,width=.45\textwidth} 
\end{tabular}
\caption{Left: the gluon momentum density from H1 and ZEUS,
together with an earlier result from NMC. The error bands are from
the experimental systematic errors. The curves are from some recent
global fits. 
Right: predictions for $F_L$ at four values of $Q^2$ from
Thorne showing the result of a standard two-loop
QCD calculation and his LORSC procedure (labelled LO(x)).}
\label{fig:gluon}
\end{figure}
$\ds {dF_2\over d\ln Q^2}\sim \alpha_SP_{gg}g(x,Q^2)$. Apart from
the global fits already mentioned both ZEUS and H1 have performed 
NLO QCD fits to extract $xg(x,Q^2)$. The advantage that the experimental
teams have is that they can include a full treatment of systematic
errors. Since HERA data does not extend to large $x$ fixed target DIS
data has to be included to fix the parameters of the valence quark
distributions. ZEUS uses its 1994 HERA data and a fixed 
$\alpha_S=0.113$, H1 fits HERA 1995/96 data and $\alpha_S=0.118$
More details are given by Prinias~\cite{prinias}. The resulting gluon
distributions are shown in Fig.~\ref{fig:gluon}(left) together with that
from the NMC experiment and some curves from global fits. The total
error is about 10\% at the lowest $x$ values. All determinations
agree within the error bands except for GRV(94) which was not fit to
the HERA(94) data and which does not describe the recent HERA data
in detail. Part of the discrepancy comes from the lower value of 
$\alpha_S$ used, but it is also known that the very low starting scale
of $0.3\,$GeV$^2$ makes the gluon distribution rise too steeply at
moderate $Q^2$ values.

Despite the manifest success of DGLAP evolution in describing $F_2$
data, the argument about low $x$ QCD continues.
If DGLAP is the full story then why are the large ln$(1/x)$ terms
suppressed? A number of authors~\cite{smallx} have investigated 
the need for including the ln$(1/x)$ terms (`resummation') 
but come to different conclusions. The most complete
approach is that of Kwiecinski, Martin and Stasto~\cite{kms} which
combines the BFKL and DGLAP equations and gives a reasonable
representation of the low $x$ data. Another approach to BFKL which
is quite successful phenomenologically is that of the colour 
dipole~\cite{royon}. Apart from the resummation of the leading
twist log terms, it has been argued recently that
higher twist (power corrections in $Q^2$) may be significant at
low $x$~\cite{bartels} and that shadowing corrections 
may be larger than BFKL effects in the kinematic region of HERA 
data~\cite{levduc}. It may be that some of diferences in outcome
can be traced to different renormalisation schemes. A way to
avoid such difficulties is to formulate the problem in terms
of {\it physical} quantities, such as $F_2$ and $F_L$, rather
than parton densities. This approach has been advocated by 
Catani~\cite{catani} and taken furthest by Thorne~\cite{thorne} 
in his Leading Order Renormalisation
Scheme Consistent (LORSC) framework. Although only at leading order
he gets slightly better fits to the low $x$ data than the 
conventional DGLAP global fits. What is crucially needed to sort
out these various ideas are measurements of another observable
as the different schemes can all fit $F_2$ but then differ for
the other. This is demonstrated in 
Fig.~\ref{fig:gluon}(right) for $F_L$.

\subsection{$F_L$ and $F_2^c$}

All fixed target experiments, except E665, have provided
measurements of $F_L$. The measurement requires collecting data at 
high $y$ for at least two centre-of-mass energies.
The most recent
measurements are from SLAC/E140X~\cite{e140x}, 
NMC~\cite{nmcf2,nmcr} and CCFR~\cite{ccfrr}. At the smallest
$x$ value of these data, $4\cdot 10^{-3}$ from NMC,
$F_L$ is possibly rising, but the errors are 
rather large. The $x$ range and precision of the $F_L$ data are
both insufficient for them to discriminate between low $x$ models. 
To date HERA has run essentially at a fixed
centre-of-mass enegry of $300\,$GeV thus precluding a direct
measurement of $F_L$.
\begin{figure}[bhp]
\vspace*{13pt}
\begin{center}
\psfig{figure=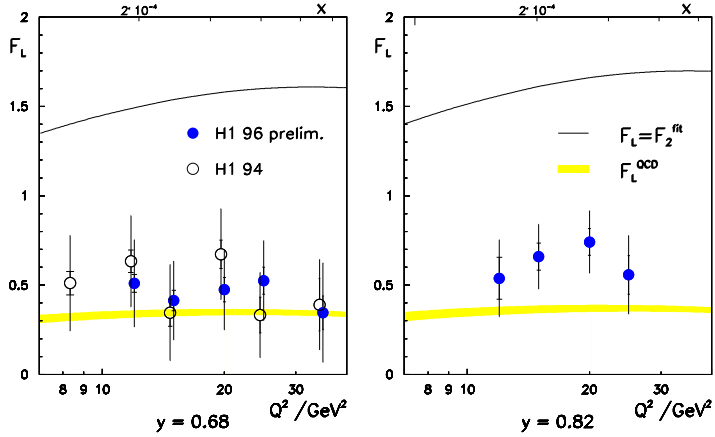,width=0.8\textwidth} 
\caption{Estimate of $F_L$ by H1 from their NLO QCD fitting
procedure as explained in the text.}
\label{fig:h1fl}
\end{center}
\end{figure}
In the meantime H1 has used NLO QCD and their high
statistics data to make an estimate of $F_L$~\cite{h1r}. The 
essence of the
idea is to determine $F_2$ for $y<0.35$ (where the contribution
of $F_L$ to the cross-section is negligible) by a NLO QCD fit.
The fit is then extrapolated to larger $y$ and used to subtract
$F_2$ from the measured cross-section. At this conference 
the results for $F_L$ were updated by preliminary data from 
the HERA 1996 run~\cite{h1_9596f2},
giving $F_L$ at $y=0.68$ and $0.82$, the results are shown in
Fig.~\ref{fig:h1fl}. The extrapolation is the
most uncertain part of the analysis. H1 has checked that using
other models for the extrapolation gives the same value for $F_L$
to within a few percent, but it has been argued that the error
could be larger~\cite{thorneh1fl}. 
The H1 estimate for $F_L$ is compatible with
pQCD calculations using recent global PDFs.
\begin{figure}[htbp]
\vspace*{13pt}
\begin{center}
\psfig{figure=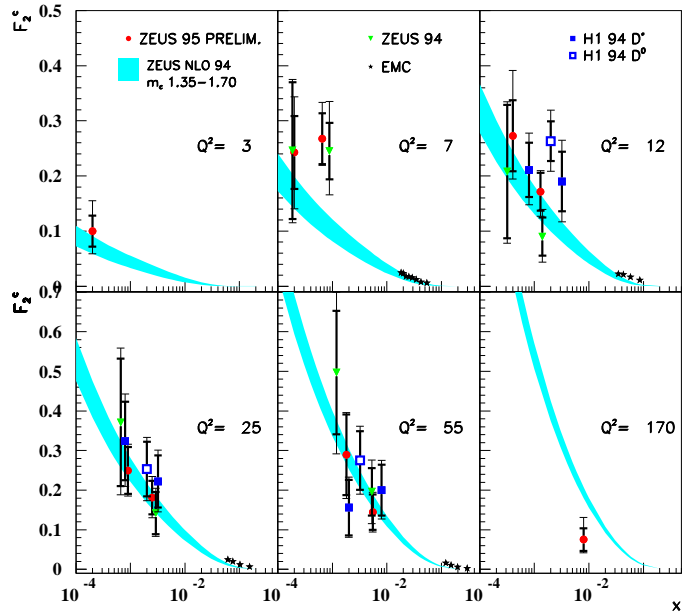,width=0.8\textwidth} 
\caption{Recent results from HERA on $F_2^c$. Earlier data from EMC is
also shown. The curve shows the result of a NLO calculation allowing 
variation of the charm mass in the range $1.35-1.7\,$GeV.}
\label{fig:charm}
\end{center}
\end{figure}

The calculation of the NLO coefficient functions for massive
quarks by Laenen et al~\cite{laenen} gave an impetus for the
question of how massive quarks should be included in NLO
global fits. The GRV(94) fit and the fits by H1 and ZEUS 
include charm only by the boson-gluon fusion (BGF) process.
It has been argued that this cannot be correct well above
threshold when the charm mass becomes negligible, charm
should then be treated as any other light quark. This 
interesting subject will not be pursued here as it can be
followed in refs~\cite{charm}, rather the status of
measurements of $F_2^c$ will be reviewed briefly. At HERA
charm can contribute up to 30\% of the cross-section, so it
is important to understand both how to describe it theoretically
and to measure it directly. The methods used by H1 and ZEUS
to tag charm are by $D^*$, $D^0$ two body decays and 
by the $D^*-D^0$ mass difference. Statistics are limited by
the small combined $D^*\to K\pi\pi$ branching ratio of only
2.6\%. A major source of systematic error is the extrapolation 
of the measured $D^*$ production cross-section to the full
phase space in rapidity and $p_T$. All these
matters are covered in more detail by Prinias~\cite{prinias}.
The HERA results for $F_2^c$ are shown in Fig.~\ref{fig:charm}
together with a NLO calculation from Harris and Smith \cite{hsmith}.
The band shows the uncertainty in the calculation, gluon densities
were taken from GRV(94) or CTEQ4F but the largest source of uncertainty
comes from the mass of the charm quark.
\begin{figure}[htbp]
\vspace*{13pt}
\begin{center}
\psfig{figure=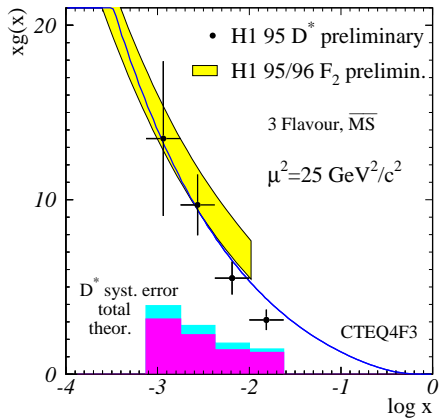,height=6.5cm} 
\caption{Extraction of the gluon density from their 1995 $D^*$
data together with the gluon density from scaling violations in
$F_2$.}
\label{fig:h1cg}
\end{center}
\end{figure}
The results are encouraging and
their precision will improve through higher luminosity and the 
use of microvertex detectors
(installed in H1, planned for ZEUS). In another
contribution~\cite{h1cg}, also covered by Prinias, H1 have used
tagged DIS charm events from HERA(95) data to make a direct 
determination of the gluon density at four $x$ values between
$0.7\cdot 10^{-3}$ and $0.5\cdot 10^{-1}$. The results are shown
in Fig.~\ref{fig:h1cg} together with the gluon density determined
by H1 from scaling violations in their 1995/6 $F_2$ data.

The material in section~\ref{sec:unpol} is covered in greater 
detail in a recent
review by Cooper-Sarkar, Devenish and De Roeck~\cite{cdd}.

\section{Nucleon Spin Structure}
\label{sec:spin}

The challenge of polarised DIS is to understand the dynamical
distribution of spin amongst the nucleon's constituents,
summarised by the relation\hfill\break
\protect{${1\over 2} = {1\over 2}\Delta\Sigma + \Delta g + 
\langle L_z\rangle$} where $\Delta\Sigma,~\Delta g$ are the
contributions of the quarks and gluons respectively and
$\langle L_z\rangle$ is the contribution from parton orbital
angular momentum. The primary measurements are the spin
asymmetries for nucleon spin parallel and
perpendicular to the longitudinally polarised lepton spin.
They are related to the polarised
structure functions $g_1,~g_2$ by kinematic factors. Only $g_1$
has a simple interpretation in terms of polarised PDFs, namely
\begin{equation}
g_1(x)={1\over 2}\sum_f e^2_f (q^\uparrow_f(x)-q^\downarrow_f(x))
={1\over 2}\sum_f e^2_f \Delta q_f(x)
\end{equation}
where the sum is over quark and antiquarks with flavour $f$ and
$q^\uparrow_f,~q^\downarrow_f$ are the quark distribution
functions with spins parallel and antiparallel 
to the nucleon spin. Full details of the formalism and QCD
evolution equations may be found in ref.~\cite{anselm}. The
observed asymmetries are reduced by the beam and target 
polarisations and the target dilution factor. Polarisations are
usually greater than 50\%, but the dilution factor is generally
quite small for solid or liquid targets, typically 0.13 for
butanol and 0.3 for $^3$He. 
\begin{table}[htb]
\caption{Summary of recent polarised structure function experiments.
}
\centerline{
\begin{tabular}{llllcccc}\\
 \hline
 Lab& Beam ~~ & Targets~~& Experiment~~  & $x$ & Status \\
 \hline
 SLAC& e 29(GeV)~~ & $^3$He, NH$_3$, ND$_3$ & E142/3  & 
 $0.03-0.8$ & complete \\
 SLAC& e 48 & $^3$He, NH$_3$, LiD & E154/5  &  
 $0.014-0.7$ & analysis \\
 CERN& $\mu$ 190  & D- H- butanol, NH$_3$~~ & SMC  &  
 $0.003-0.7$ & complete \\
 DESY& e 27 & H, $^3$He & HERMES  & 
 $0.023-0.6$ & running \\
\hline\\
\end{tabular}}
\label{tab:spin}
\end{table}
The HERMES experiment
at DESY uses a polarised internal gas jet target in the HERA-e
beam and thus achieves a dilution factor of 1. Apart from HERMES,
the latest round of experiments from SLAC and CERN is almost complete,
details are given in Table~\ref{tab:spin}. From the table
it can be seen that the measurements
of polarised DIS do not reach very small values of $x$, the largest
range is that of the SMC experiment.

\begin{figure}[htb]
\vspace*{13pt}
\begin{tabular}[ht]{ll}
\psfig{figure=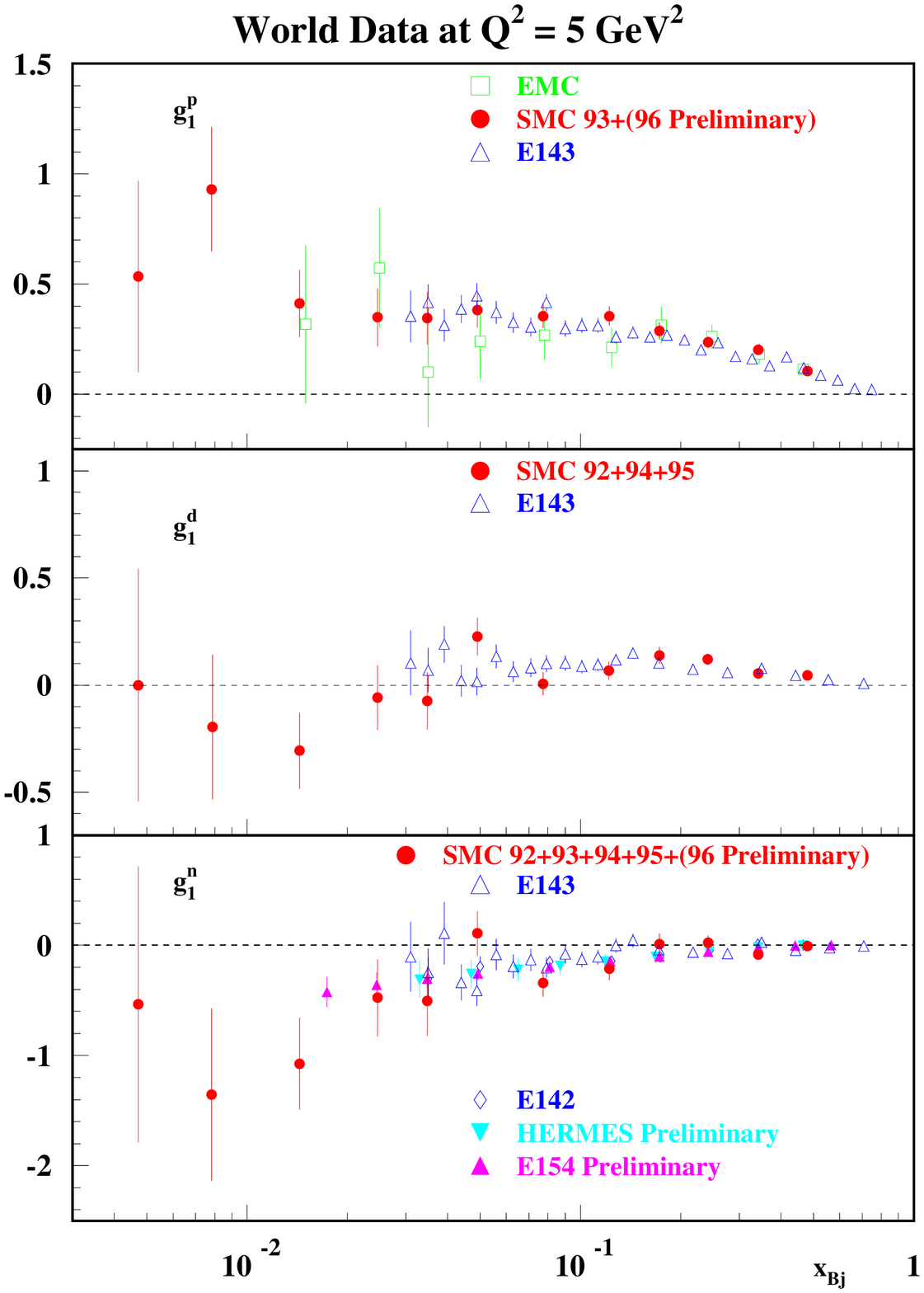,width=.45\textwidth}&
\psfig{figure=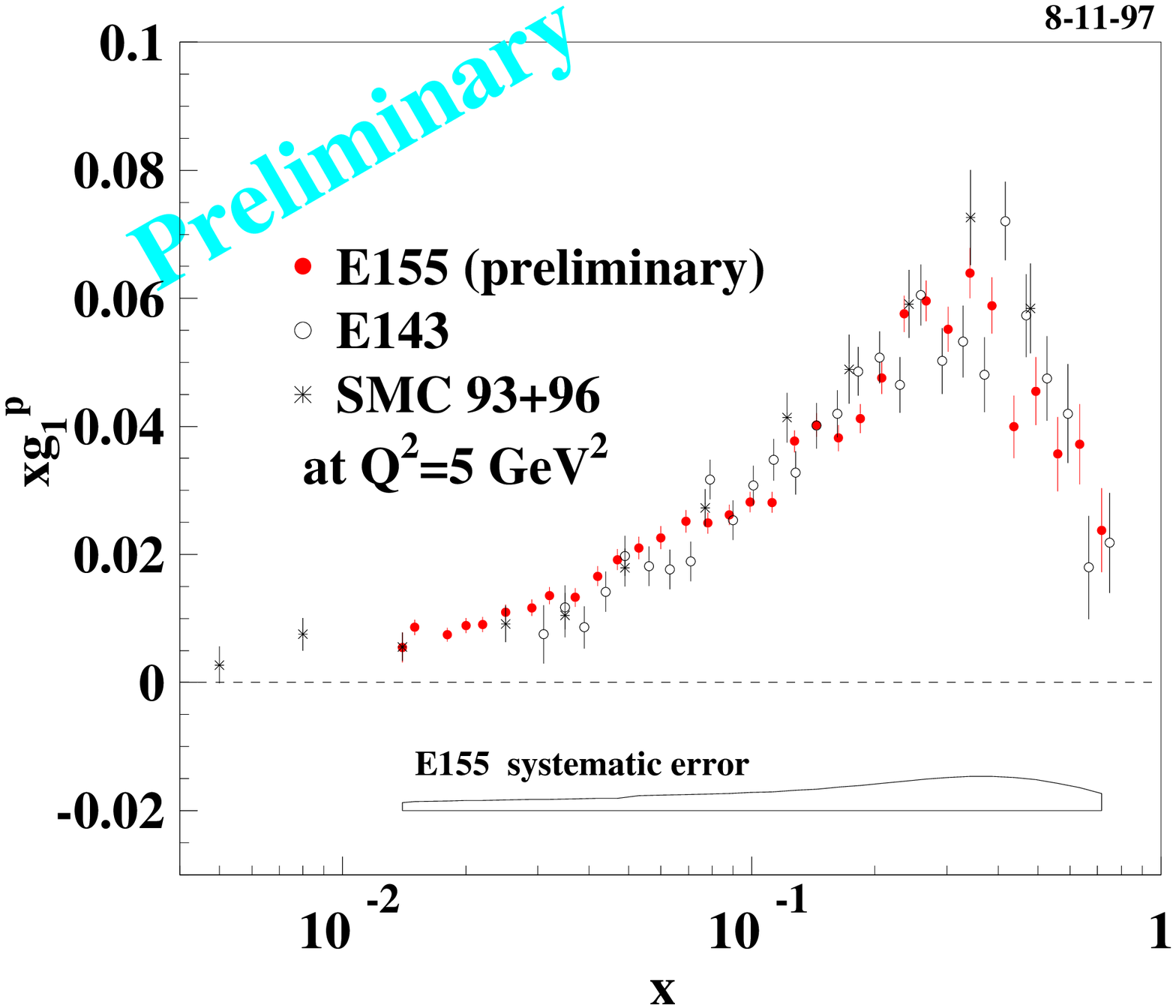,width=.45\textwidth} 
\end{tabular} 
\caption{Left: a compilation by the SMC collaboration of data on
$g_1$ for protons, neutrons and deuterons from SMC, SLAC and
HERMES polarised DIS experiments.
Right: Preliminary data for $xg_1^p$ from the SLAC E155 polarised 
DIS experiment, compared to earlier data.}
\label{fig:allg1}
\end{figure}
Most of the data is for $g_1$
and there is nice agreement between the different experiments as
can be seen from Fig.~\ref{fig:allg1}(left). There is a small amount of data
for $g_2$ from the SLAC experiments, $g_2$ is small and consistent both
with zero and the expectation of the twist-2 calculation.
More details on the individual experiments are to be found in the
contributions of Souder~\cite{souder} (E154/5), Le Goff~\cite{legoff}
(SMC) and Blouw~\cite{blouw} (HERMES). New preliminary data on the 
proton asymmetry comes from the HERMES collaboration~\cite{blouw} 
and the SLAC E155 experiment~\cite{e155}, both offer the prospect
of reduced statistical errors as can be seen for E155 from 
Fig.~\ref{fig:allg1}(right).

\begin{figure}[hbt]
\vspace*{13pt}
\begin{tabular}[ht]{ll}
\psfig{figure=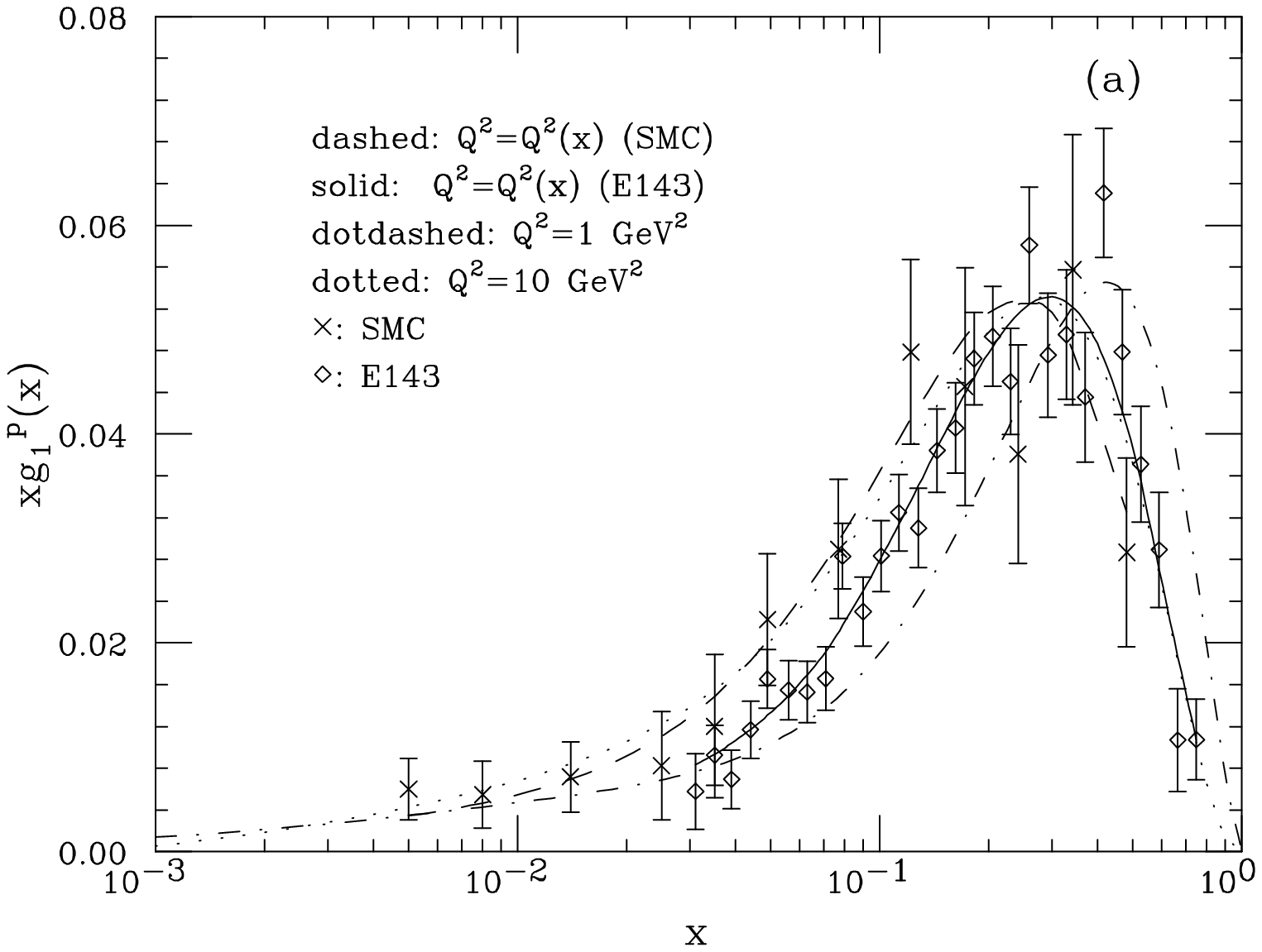,width=.45\textwidth}&
\psfig{figure=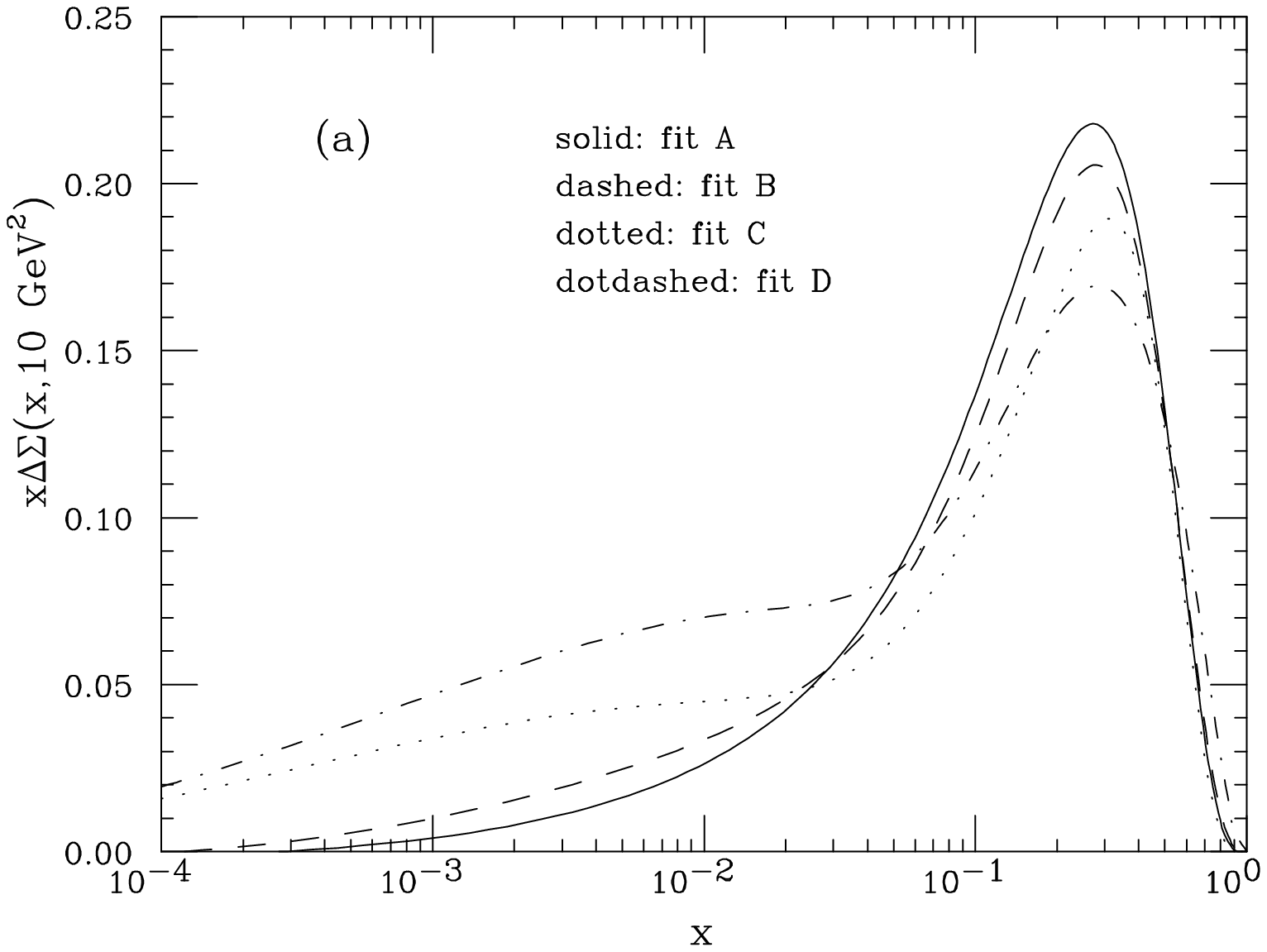,width=.45\textwidth} \\
\psfig{figure=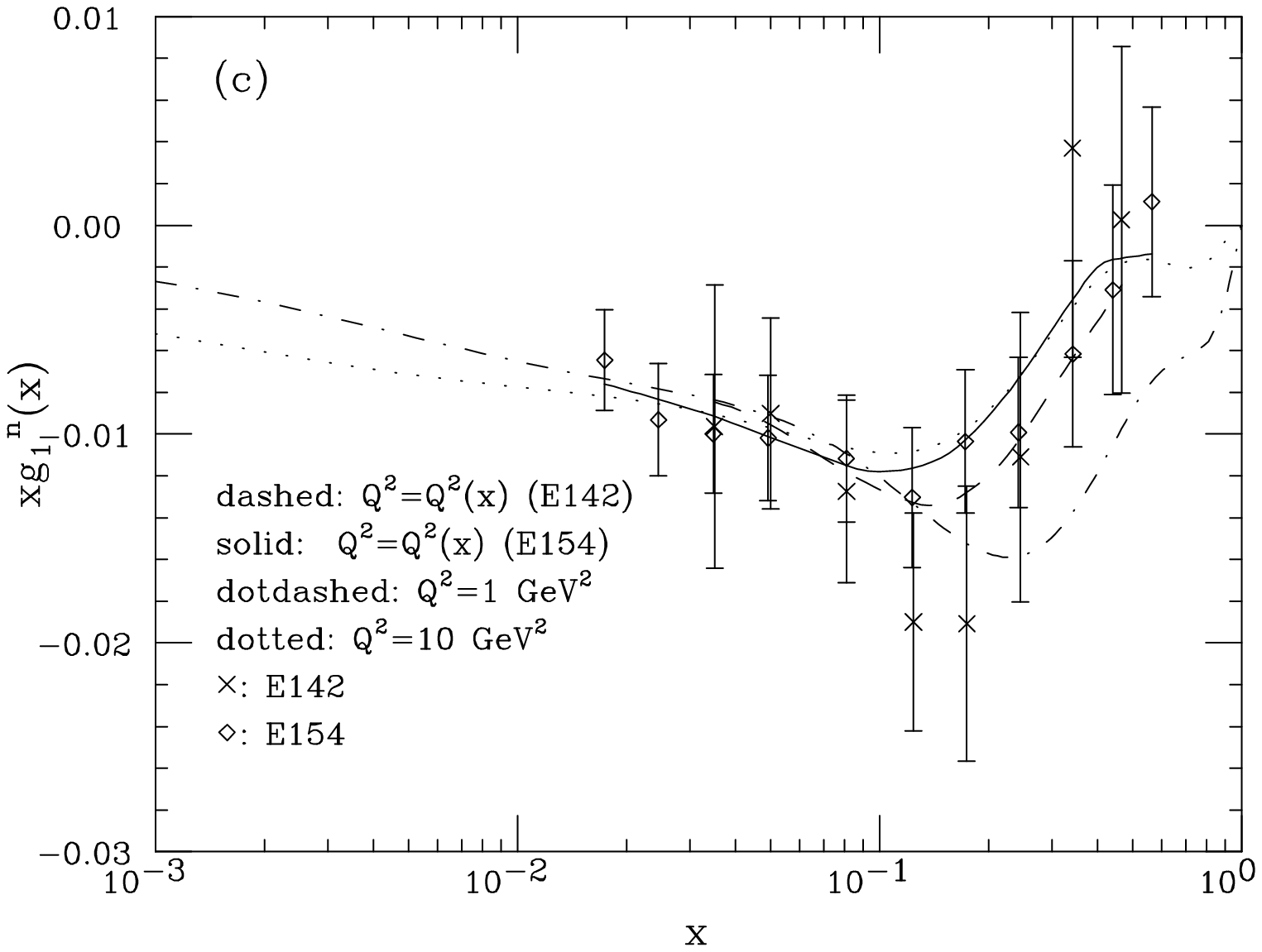,width=.45\textwidth}&
\psfig{figure=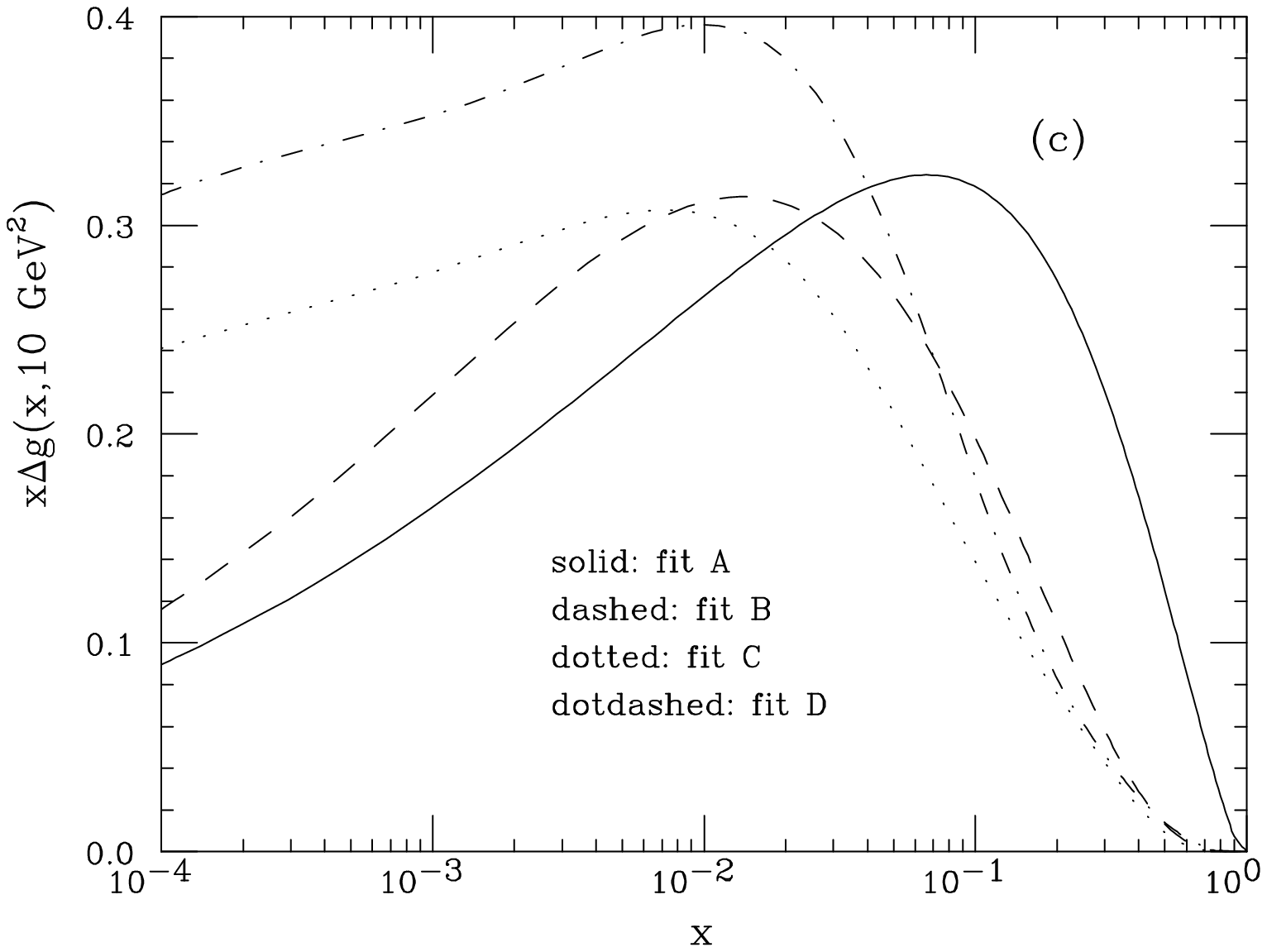,width=.45\textwidth} 
\end{tabular}
\caption{From the ABFR NLO QCD fit to polarised DIS data. Left (a)
$xg_1^p$, (c) $xg_1^n$; right (a) $x\Delta \Sigma$, (c) $x\Delta g$.
Curves A, B, C, D refer to fits with different assumptions for
low $x$ behaviour.}
\label{fig:abfr}
\end{figure}

Apart from more accurate data, the big advance this year
has been the extensive use of NLO QCD fitting. Apart from the intrinsic
interest in testing QCD, the NLO fit also gives the best
extrapolation of the data to a common $Q^2$ for the
evaluation of sum rum integrals $\ds \Gamma_1^i(Q^2)=\int_0^1g_1^i(x,Q^2)dx$
and the evaluation of separate parton components.
The first NLO fits were performed in 1995/6~\cite{nlo956}, this year
the experimental groups SLAC/E154~\cite{souder} and SMC~\cite{legoff} 
and the theoretical teams of Altarelli et al (ABFR)~\cite{abfr} and
Leader et al (LSS)~\cite{lss} have published such analyses. There are
considerable differences of detail in the approaches taken by the 
different groups, perhaps the most important is the choice of 
factorisation scheme, LSS use $\overline{\rm MS}$ and all other groups
follow the Adler-Bardeen scheme to give a scale independent
first moment for $\Delta \Sigma$,
$\Delta \Sigma_{AB}=\Delta q_0+n_f{\alpha_S\over 2\pi}\Delta g$.
All groups assume a non-singular $x$ dependence for the input 
distributions at $Q^2_0$ and the partonic constraint 
$|\Delta q_{NS}|<q_{NS}$, where $NS$ refers to the non-singlet contribution.
The quality of the fits is good and one finds that the non-singlet
valence quark distributions are quite well determined. The
results for the quark singlet and gluon distributions are less good
as there are no data for $x<3\cdot 10^{-3}$. These features are
shown in Fig.~\ref{fig:abfr} from the ABFR fits, the two left hand
plots show the quality of the fit to data (fit B) and the two right
plots show $\Delta \Sigma$ (upper) and $\Delta g$ (lower) for a
variety of different assumptions about the low $x$ behaviour 
(see~\cite{abfr} for details). 

In addition to extrapolation in $Q^2$, to evaluate $\Gamma_1^i$
the data must also be extrapolated in $x$. There is no problem 
as $x\to 1$, but
there is still considerable uncertainty as $x\to 0$. This is of 
course a reflection of both the lack of data and the range of
possible behaviours for the singlet distributions at small $x$.
SMC has investigated this point in some detail~\cite{legoff}
for the evaluation of $\Gamma_1^p$. Generally  $\Gamma_1^p$ is
measured to about 10\% and  $\Gamma_1^n$ to about 20\%. For the
parton components, the quark integral is known to about 10\% but
the gluon integral only to 40\% (ABFR) and more like 100\% error
from the experimenters fits. All agree that the gluon contribution
is positive. 

\begin{figure}[htbp]
\vspace*{13pt}
\begin{center}
\psfig{figure=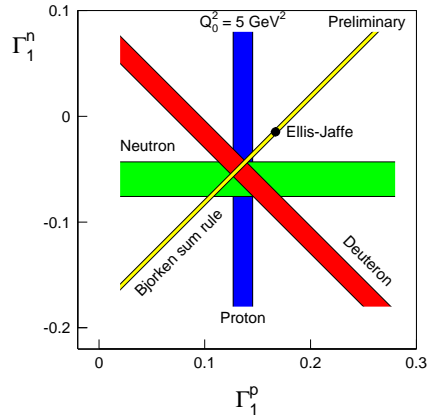,height=6cm} 
\caption{Summary of the status of experimental determination
of the integrals $\Gamma_1^p,~\Gamma_1^d,~\Gamma_1^d$ and the
Bjorken and Ellis-Jaffe sum rules by the SMC collaboration.}
\label{fig:smcsr}
\end{center}
\end{figure}

What does this mean for the sum rules?
The fundamental Bjorken sum rule $\Gamma_1^p-\Gamma_1^n=
{C_1^{NS}(Q^2)\over 6}\left|{g_A\over g_V}\right|$, where $C_1^{NS}$
is a QCD coefficient known to order $\alpha_S^3$, is found to
be reasonably well satisfied, at about the 10\% level, by all groups.
The theoretically less well found
Ellis-Jaffe sum rules for $\Gamma_1^p$, $\Gamma_1^n$ separately
are violated at the $2\sigma$ level. The overall situation is 
summarised in Fig.~\ref{fig:smcsr}.

For inclusive measurements, this situation will not improve until
there is data at smaller values of $x$, from RHIC or
a polarised HERA collider. Another way to learn about individual
parton distributions is through the measurement of semi-inclusive
asymmetries. This type of
measurement has been pioneered by the SMC whose latest
results are reported by Baum~\cite{baum}. HERMES has also
presented some preliminary semi-inclusive results to this 
conference \cite{blouw}. Asymmetries for identified
particles such as positive or negative hadrons have been measured
and from these the valence contributions $\Delta u_V$, $\Delta d_V$
and the sea quark $\Delta \overline{q}$ (with some additional
assumptions) determined. In the future such techniques applied
to charmed particles will help to pin down the gluon contribution
as well.

\section{$F_2^{\gamma}$}
\label{sec:f2g}

The photon structure functions both for the two-lepton
final states and the hadronic final state, $F_2^\gamma$, have
been measured from two-photon interactions at LEP. The
details of the measurements by ALEPH, DELPHI and OPAL
are covered in the mini-review talk by Nisius~\cite{nisius}.
In such measurements, only one scattered $e^\pm$ is detected (to give
the $Q^2$ of the event), the other giving the `target' photon
is lost down the beam pipe. $Q^2$ is measured
almost directly from the tagged lepton, but $x$ has to be
deduced from the measured final state particles. The Monte Carlo
modelling of the physics and detectors is thus very important
and some significant improvements have been made in these areas
recently~\cite{miller}. The increase of the beam energies
in LEP2 operations has increased both the phase space and the
statistics for two-photon physics. 
\begin{figure}[htbp]
\vspace*{13pt}
\begin{center}
\psfig{figure=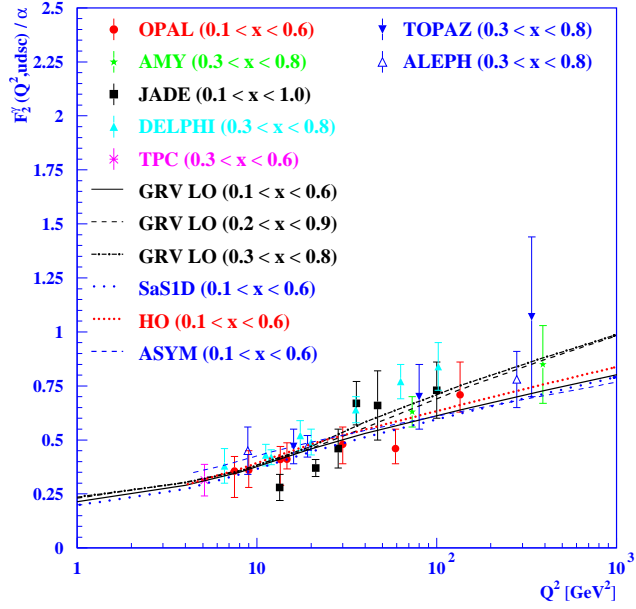,height=8cm} 
\caption{Summary plot showing $F_2^\gamma$ at medium $x$ 
as a function of $Q^2$ compared
to various LO and NLO QCD calculations as labelled.}
\label{fig:ggf2q2}
\end{center}
\end{figure}

Apart from the interest in
determining the partonic content of the photon, the QCD evolution
equations involve an inhomogeneous term which can be calculated
from $\gamma \to q\overline{q}$ splitting.
The data for $F_2^\gamma$
are well described by NLO QCD fits and the larger lever arm
in $Q^2$ allows one to see for the first time the logarithmic
increase of  $F_2^\gamma$ with $Q^2$, as demonstrated in
Fig.~\ref{fig:ggf2q2}.
Because of the limited reach in small $x$ at LEP, it has not been
possible to determine if $F_2^\gamma$ rises steeply as $x$ 
decreases. For the same reason the gluon component of
the NLO fits is not well determined. 

\begin{figure}[htbp]
\vspace*{13pt}
\begin{tabular}[ht]{ll}
\psfig{figure=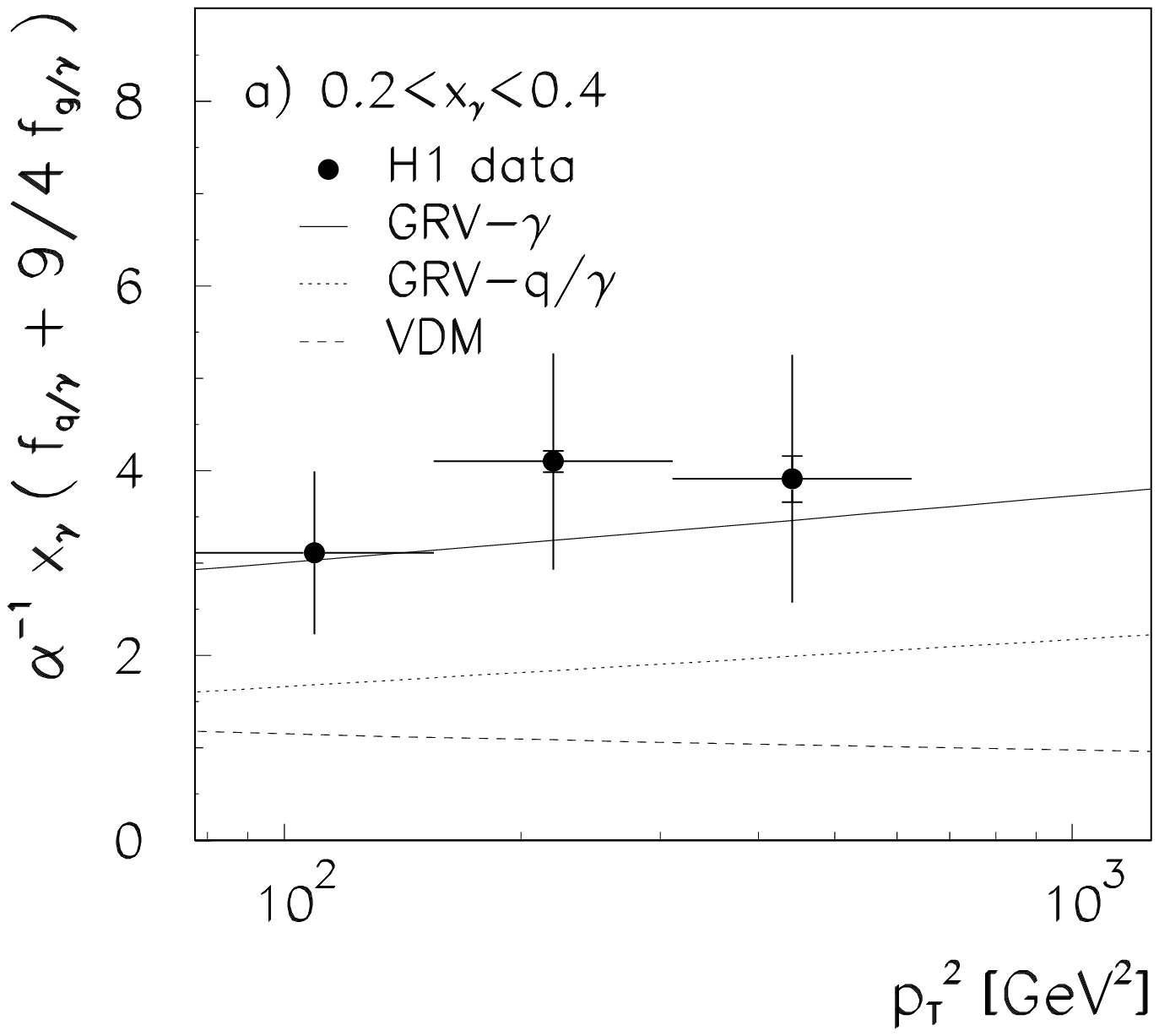,width=.45\textwidth}&
\psfig{figure=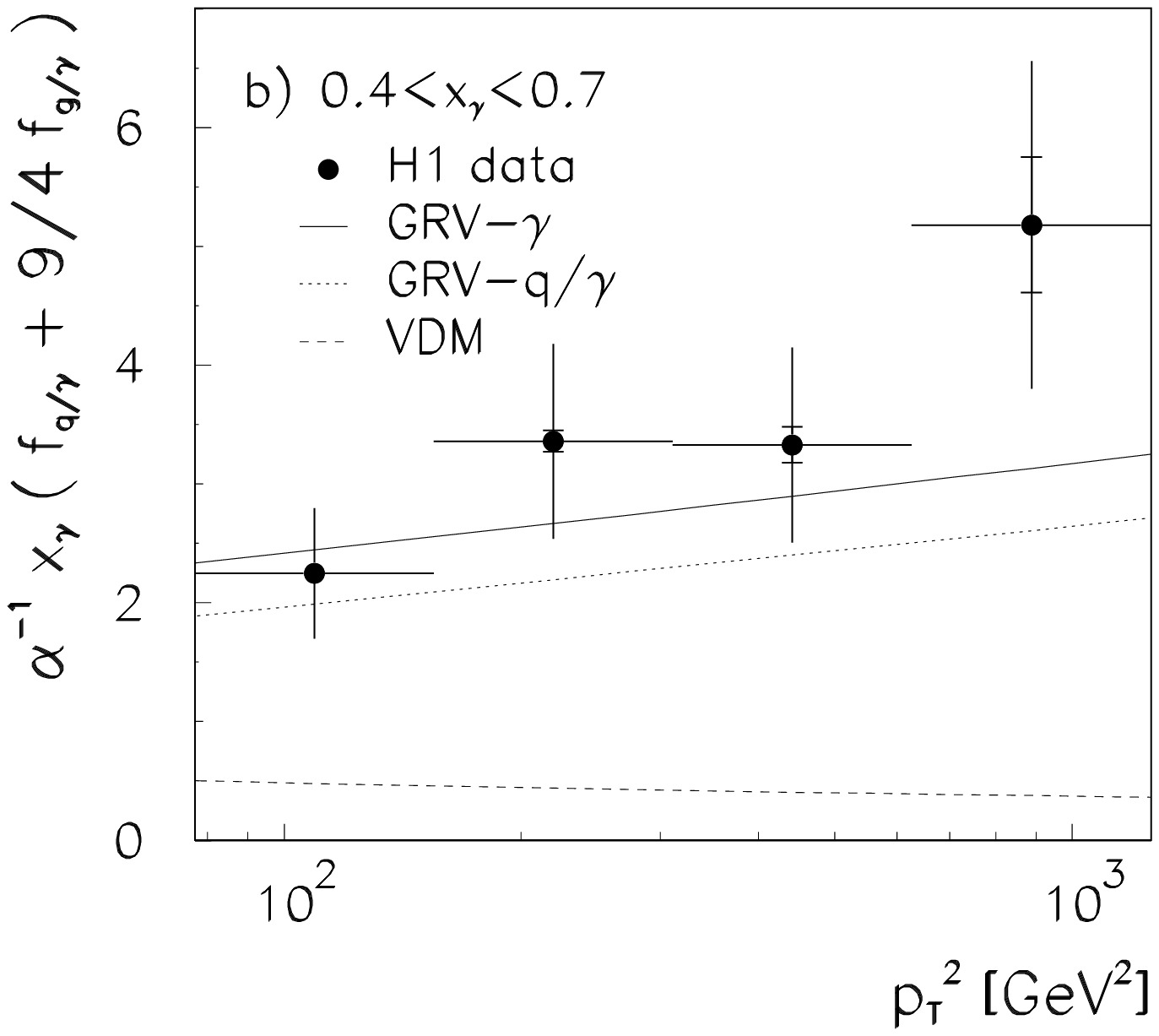,width=.45\textwidth} 
\end{tabular} 
\caption{The effective photon parton density $\overline{f_\gamma}$ extracted
from H1 1994 dijet data. The curves were calculated using GRV-LO partons
and are the complete calculation (full curve), the quark component only
(dotted) and the vector meson part (dashed).}
\label{fig:h1pdfg}
\end{figure}

Photoproduction processes at HERA also give information on
photon structure. The process $\gamma p \to j_1j_2X$ is 
particularly attractive as it is sensitive to both direct and 
resolved photon processes and the kinematic variables $x_\gamma$ 
and $p_t^2$ (equivalent to $Q^2$) can be reconstructed from the 
final state jets. H1 have used used such a measurement to extract an
effective  photon PDF
$\overline{f_\gamma}=f_{q/\gamma}+{9\over 4}f_{g/\gamma}$ for
$0.2<x_\gamma<0.7$, the results are shown in Fig.~\ref{fig:h1pdfg}.
More details are given in the talk by Muller~\cite{muller}.

Finally a more phenomenological approach to the problem of
the small $x$ region is described by 
Gurvich~\cite{gurvich} in which $F_2^p$ data at low $x$ is
used with Gribov factorisation to generate low $x$ $F_2^\gamma$
`data'. The generated and directly measured $F_2^\gamma$ data
are then fit using leading order QCD evolution over the range
$4.3<Q^2<390\,$GeV$^2$.

\section{Summary and Outlook}

Generally the measurements of structure functions are in good shape
and the data are well described by NLO QCD.
For $F_2$ at low $x$ more work is needed to understand fully the 
implications for QCD and accurate data for an other observable such
as $F_L$ or $F_2^c$ are essential. The understanding of
both polarised structure functions and $F_2^\gamma$ 
at small $x$ is hampered by lack of data. 
Information on the nucleon gluon density is being provided by the use 
of charm tagging and this will improve.

For the future we can look forward to the completion of the two-photon
programme at LEP2 and the large increase in luminosity promised by
the HERA upgrade. The COMPASS experiment at CERN, polarised scattering
at RHIC and maybe a fully polarised HERA hold out the promise of
finally unravelling the mysteries of nucleon spin. 

The measurement and analysis of DIS and structure functions are
still challenging our understanding of hadronic structure and QCD 30 years
after the discovery of scaling.

\begin{flushleft}
{\bf Acknowledgements}
\end{flushleft}
I thank my colleagues in H1 and ZEUS for providing information and
insights on the latest HERA results. I thank A. Br\"ull, J. Le Goff
and E. Rondio for instructions on spin physics, likewise
D. Miller, R. Nisius and S. S\"oldner-Rembold on two-photon physics.
I thank the organisers of the Conference, particularly D. Lellouch,
and IT staff at Oxford and Jerusalem for much practical help.

% ---- Bibliography ----
%

\end{document}